\documentclass[conference,a4paper]{IEEEtran}
\IEEEoverridecommandlockouts
\usepackage[utf8]{inputenc}
\usepackage[T1]{fontenc}
\usepackage{url}              
\usepackage{cite}             

\usepackage[cmex10]{amsmath}  
\usepackage{amssymb}
\usepackage{bm}
\usepackage{mleftright}       
\mleftright                   

\usepackage{graphicx}         
\usepackage{subfigure}

\usepackage{booktabs}         
\usepackage{color}
\usepackage{marvosym}
\usepackage{ntheorem}
\usepackage{relsize}

\newtheorem{proposition}{\underline{Proposition}}

\begin{document}
	\title{MIMO Radar Transmit Signal Optimization for Target Localization Exploiting Prior Information}
	\author{
		\IEEEauthorblockN{Chan Xu and Shuowen Zhang}
		\IEEEauthorblockA{Department of Electronic and Information Engineering, The Hong Kong Polytechnic University\\
			Email: \{chan.xu,shuowen.zhang\}@polyu.edu.hk}\thanks{This work was supported in part by the General Research Fund from the Hong Kong Research Grants Council under Grant 15230022, and in part by the National Natural Science Foundation of China under Grant 62101474.}}
	
	\maketitle

	\begin{abstract}
		In this paper, we consider a multiple-input multiple-output (MIMO) radar system for localizing a target based on its reflected echo signals. Specifically, we aim to estimate the \emph{random} and \emph{unknown} angle information of the target, by exploiting its prior distribution information. First, we characterize the estimation performance by deriving the \emph{posterior Cram\'er-Rao bound (PCRB)}, which quantifies a lower bound of the estimation mean-squared error (MSE). Since the PCRB is in a complicated form, we derive a tight upper bound of it to approximate the estimation performance. Based on this, we analytically show that by exploiting the prior distribution information, the PCRB is always no larger than the Cram\'er-Rao bound (CRB) averaged over random angle realizations without prior information exploitation. Next, we formulate the transmit signal optimization problem to minimize the PCRB upper bound. We show that the optimal sample covariance matrix has a rank-one structure, and derive the optimal signal solution in \emph{closed form}. Numerical results show that our proposed design achieves significantly improved PCRB performance compared to various benchmark schemes.
	\end{abstract}
	\section{Introduction}
	Multiple-input multiple-output (MIMO) radar can enhance the localization performance via exploiting the waveform diversity \cite{1}, thus has attracted significant research attention over the years. Specifically, a MIMO radar can transmit noncoherent known signals and receive the reflected signals (echoes) via multiple antennas. The opportunity to harvest waveform diversity offers high resolution and sensitivity, good parameter identifiability, and direct applicability of adaptive array techniques \cite{2}. To take full advantage of the degrees-of-freedom (DoFs) brought by multiple antennas, transmit signal design is of paramount importance for MIMO radar systems.
	
	Generally speaking, the existing literature on MIMO radar transmit signal design can be categorized into two classes: beampattern approximation and direct design for localization performance optimization. In the first class, the transmit signals are designed to approximate a desired and pre-designed beampattern \cite{3,4,5}. The localization performance is implicitly reflected by the difference between the desired beampattern and the approximated beampattern, which cannot be explicitly quantified. On the other hand, in the second class, the mean-squared error (MSE) is a commonly adopted metric to assess the performance of localization. However, since the minimum possible MSE is generally difficult to characterize, some lower bounds of the MSE have been proposed, among which the most well-known one is the \emph{Cram\'er-Rao bound (CRB)} \cite{6}. For MIMO radar systems, the expressions of CRB for angle estimation \cite{7} and velocity estimation \cite{8} have been derived. With CRB as the performance metric, various works have studied the transmit signal optimization, e.g., \cite{9,10}.
	\begin{figure}[t]
		\centering
		\includegraphics[width=2.8in]{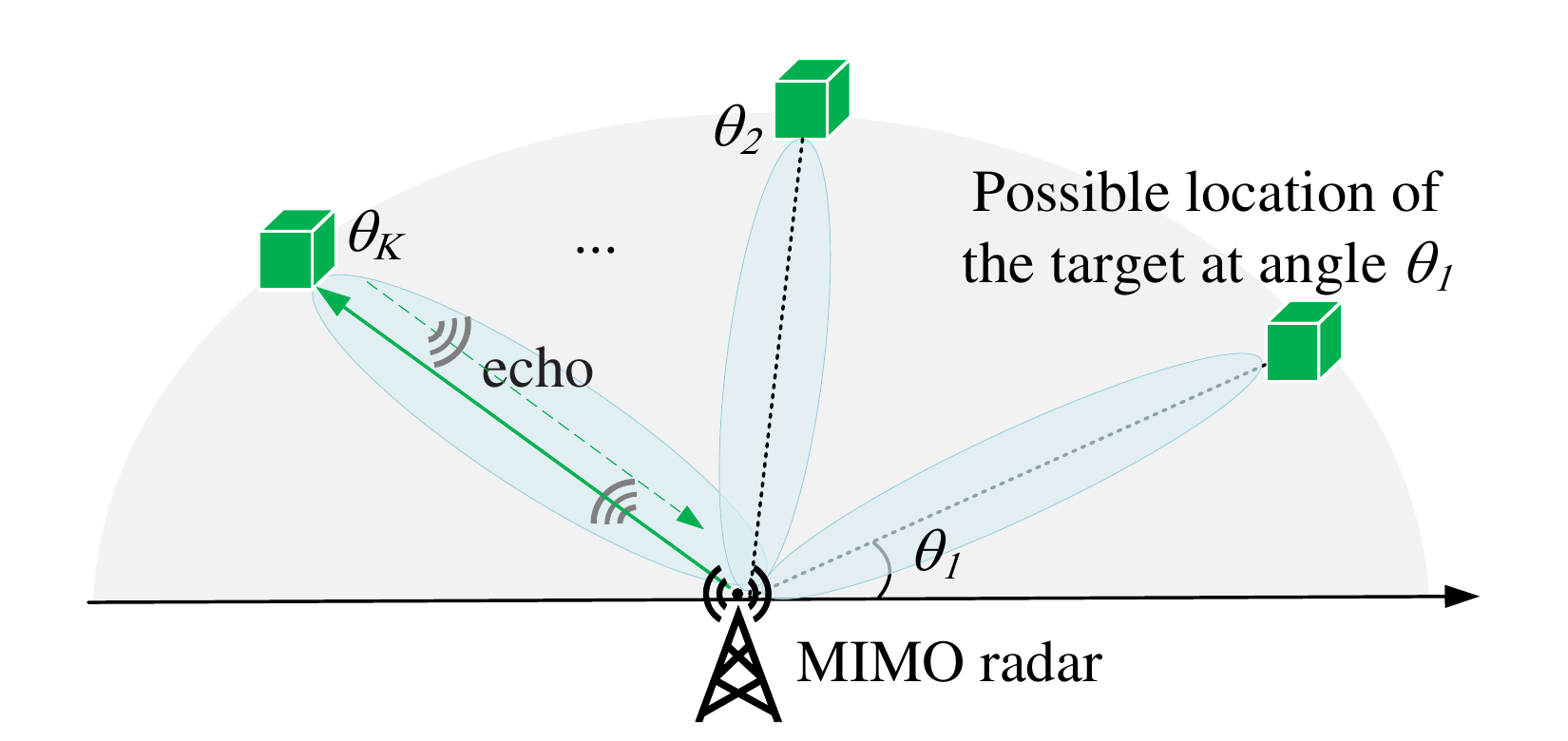}
		\vspace{-4mm}
		\caption{Illustration of target localization via a MIMO radar system.}\label{fig_sys}
		\vspace{-7mm}
	\end{figure}
	
	The vast majority of the existing literature focused on the case where the location parameters to be estimated are \emph{deterministic}. However, in practice, the location parameters can be \emph{random}, for which the distributions can be known \emph{a priori}. For example, for a mobile vehicle or pedestrian, the location parameters at the upcoming time slots are generally functions of the location parameters at the current and previous time slots, for which the distributions can be obtained based on the previous localization results and/or by exploiting empirical data. With the prior information exploited, a so-called \emph{posterior Cram\'er-Rao bound (PCRB)} can be derived to characterize the lower bound of MSE \cite{van}. Nevertheless, to the best of our knowledge, how to optimize the MIMO radar transmit signals for optimizing the estimation performance of random parameters by exploiting its prior distribution information is still an open problem, which motivates our study in this paper.
	
	This paper studies the target localization via a MIMO radar system with co-located transmit and receive antennas. The angular location of the target is modeled as a random variable, which is estimated via the signals sent from the MIMO radar transmitter, reflected by the target, and received back at the MIMO radar receiver. To characterize the angle estimation performance exploiting its prior distribution information, we first characterize the PCRB of the MSE, which is in a complicated form. We then derive a tractable and tight upper bound for the PCRB. Based on this, we analytically show that by exploiting prior distribution information, the PCRB is always no larger than the average CRB without exploiting the prior information. Then, we formulate an optimization problem for the sample covariance matrix of the transmit signal, with the objective of minimizing the PCRB upper bound. The optimal solution to the problem is derived in closed form. Finally, numerical results verify the tightness of the proposed PCRB upper bound and validate the performance gain of the proposed transmit signal design over various benchmark schemes.

	\section{System model}\label{sec_sys}	
	We consider a MIMO radar system with $N_t\geq 1$ transmit antennas and $N_r\geq 1$ co-located receive antennas. We aim to estimate the \emph{unknown} and \emph{random} location parameter of a point target via exploiting the prior information of its distribution. Specifically, we consider a two-dimensional (2D) polar coordinate system with the reference point of the MIMO radar system being the origin, as illustrated in Fig. \ref{fig_sys}. For the purpose of drawing essential insights, we assume that every possible target location has the same distance $r\geq 0$ in meters (m) and a different angle with respect to the origin, where the common distance (range) information $r$ is known \emph{a priori}.\footnote{The range information can be either known as prior information of the target, or estimated efficiently via e.g., time-of-arrival (ToA) methods.} Thus, the only unknown and random parameter is the target's angle denoted by $\theta\in [0,2\pi)$. Motivated by practical scenarios where the target's angle distribution is typically concentrated around one or multiple nominal angles, we assume that the probability density function (PDF) of $\theta$ follows a Gaussian mixture model, which is the weighted summation of $K\geq 1$ Gaussian PDFs, with each $k$-th Gaussian PDF having mean $\theta_k\in [0,2\pi)$, variance $\sigma_k^2$,\footnote{We consider $\sigma_k^2$'s that are sufficiently small such that the probability for $\theta$ under the Gaussian mixture model to exceed the $[0,2\pi)$ region is negligible.} and carrying a weight of $p_k\in [0,1]$ that satisfies $\sum_{k=1}^K p_k=1$. Hence, the PDF of $\theta$ is given by
	\begin{align}\label{angle}
		p_\Theta(\theta)=\sum_{k=1}^{K}p_k\frac{1}{\sqrt{2\pi}\sigma_k} e^{-\frac{(\theta-\theta_k)^2}{2\sigma_k^2}}.
	\end{align}
	Note that the considered Gaussian mixture model can characterize a wide range of practical scenarios by choosing different parameters. For example, when $K$ is sufficiently large, the PDF will tend to be uniform; while when $K=1$, the PDF will reduce to the Gaussian PDF.
	
	To estimate the unknown random location parameter $\theta$, the MIMO radar transmitter sends a sequence of probing signals, which will be reflected by the target back to the MIMO radar receiver; $\theta$ is then estimated by processing the received echo signals. We consider a line-of-sight (LoS) propagation environment where no obstruction/scatter exists between the MIMO radar tranceiver and each possible target location. The overall channel from the MIMO radar transmitter to the MIMO radar receiver via target reflection is given by
	\begin{align}
		\bm{G}(\theta)=\bm{h}_R(\theta)\psi\bm{h}^H_T(\theta).	
	\end{align}
	Specifically, $\psi\in \mathbb{C}$ denotes the radar cross-section (RCS) coefficient, which is an unknown and deterministic parameter. $\bm{h}_R(\theta)=\frac{\sqrt{\beta_0}}{r}\bm{b}(\theta)$ and $\bm{h}^H_T(\theta)=\frac{\sqrt{\beta_0}}{r}\bm{a}^H(\theta)$ denote the target-receiver and transmitter-target channel vectors, respectively, where $\beta_0$ denotes the reference channel power at reference distance $1$ m; $\bm{a}^H(\theta)$ and $\bm{b}(\theta)$ denote the transmit/receive antenna array steering vectors given by  $a_i(\theta)=e^{\frac{-j\pi d(N_t-2i+1)\sin\theta}{\lambda}},\ i=1,...,N_t$, $b_m(\theta)=e^{\frac{-j\pi d(N_r-2m+1)\sin\theta}{\lambda}},\ m=1,...,N_r$, with $d$ denoting the antenna spacing in m and $\lambda$ denoting the wavelength in m. For simplicity, we define $\alpha\overset{\Delta}{=}\frac{\beta_0}{r^2}\psi=\alpha_R+j\alpha_I$ as the overall reflection gain containing both the two-way channel gain and RCS, which yields $\bm{G}(\theta)=\alpha \bm{b}(\theta)\bm{a}^H(\theta)$. It is worth noting that in general, $\alpha$ is an unknown \emph{deterministic} parameter.
	
	Let $L\geq 1$ denote the number of samples of the transmit probing signal used for the estimation of $\theta$. Let $\bm{x}_l\in \mathbb{C}^{N_t\times 1}$ denote the baseband equivalent probing signal vector at the $l$-th sample. The collection of probing signals over $L$ samples is denoted by $\bm{X}=[\bm{x}_1,...,\bm{x}_L]$, for which $\bm{R}_X=\frac{1}{L}\sum_{l=1}^L\bm{x}_l\bm{x}_l^H =\frac{1}{L}\bm{XX}^H$ denotes the sample covariance matrix. Let $P$ denote the total power constraint among all the MIMO radar transmit antennas, which yields $\mathrm{tr}(\bm{R}_X)\leq P$. The received signal vector at the $l$-th sample is given by
	\begin{align}
		\bm{y}_l=\bm{G}(\theta)\bm{x}_l+\bm{n}_l,\ l=1,...,L,
	\end{align}
	where $\bm{n}_l\sim\mathcal{CN}(0,\sigma^2\bm{I}_{N_r})$  denotes the circularly symmetric complex Gaussian (CSCG) noise at the MIMO radar receive antennas, with $\sigma^2$ denoting the average noise power. The collection of received signal vectors over $L$ samples is thus given by
	\begin{align}\label{Y}
		\bm{Y}=[\bm{y}_1,...,\bm{y}_L]=\bm{G}(\theta)\bm{X}+[\bm{n}_1,...,\bm{n}_L].
	\end{align}
	
	Note that the received signals in $\bm{Y}$ and consequently the performance of estimating $\theta$ are critically determined by the MIMO radar transmit signal design, particularly for the case considered in this paper where prior distribution information about $\theta$ is available for exploitation. For example, the transmit signals should be designed such that the radiated signal power is more concentrated over the possible target angles with high probabilities, to optimally utilize the available transmit power. In this paper, we will first characterize the performance of estimating $\theta$ by exploiting the prior distribution information, based on which we will then investigate the optimization of the MIMO radar transmit signals.
	
	\section{Estimation Performance Characterization via PCRB}\label{sec_PCRB}
	Conventionally, CRB has been widely adopted to characterize the estimation performance of unknown deterministic parameters, which is a lower bound of the MSE. In this section, by exploiting the prior distribution information of the unknown random parameter $\theta$, i.e., $p_\Theta(\theta)$ in (\ref{angle}), we propose to derive the PCRB of the MSE as the estimation performance metric.
	
	\subsection{Derivation of PCRB}
	We aim to estimate $\theta$ from the collection of MIMO radar received signals $\bm{Y}$, which is a function of both the unknown random parameter $\theta$ and the unknown deterministic parameter $\alpha$. Hence, $\alpha$ needs to be jointly estimated with $\theta$ to obtain an accurate estimation of $\theta$. For ease of exposition, we define $\bm{\zeta}=[\theta,\alpha_R,\alpha_I]^T$ as the collection of all the unknown parameters.
	
	The joint distribution of the observation $\bm{Y}$ and unknown parameter $\bm{\zeta}$ can be expressed as
	\begin{equation}\label{jointpdf}
		f(\bm{Y},\bm{\zeta})=f(\bm{Y}|\bm{\zeta})p_Z(\bm{\zeta}),
	\end{equation}
	where $f(\bm{Y}|\boldsymbol{\zeta})$ denotes the conditional PDF of $\bm{Y}$ given $\bm{\zeta}$; $p_Z(\bm{\zeta})$ denotes the marginal distribution of $\boldsymbol{\zeta}$.
	
	Note that since $\boldsymbol{\zeta}$ consists of a random parameter $\theta$ for which the distribution is known, the information of $\boldsymbol{\zeta}$ can be extracted by jointly exploiting the conditional PDF $f(\bm{Y}|\boldsymbol{\zeta})$ of the observation $\bm{Y}$ and the prior information of $\theta$. Specifically, based on (\ref{jointpdf}), the Fisher information matrix (FIM) for estimating $\boldsymbol{\zeta}$ is given by \cite{shen}:
	\begin{equation}\label{FIM}
		\bm{F} = \bm{F}_o +\bm{F}_p,
	\end{equation}
	where $\bm{F}_o$ represents the FIM from observation given as
	\begin{equation}\label{Fo}
		\bm{F}_o=\mathbb{E}_{\bm{Y},\boldsymbol{\zeta} }\left[\frac{\partial \ln(f(\bm{Y}|\boldsymbol{\zeta}))}{\partial \boldsymbol{\zeta} }\left(\frac{\partial \ln(f(\bm{Y}|\boldsymbol{\zeta}))}{\partial \boldsymbol{\zeta} }\right)^H\right];
	\end{equation}
	$\bm{F}_p$ represents the FIM from prior information given as
	\begin{equation}\label{Fp}
		\bm{F}_p=\mathbb{E}_{\boldsymbol{\zeta} }\left[\frac{\partial \ln(p_Z(\boldsymbol{\zeta} ))}{\partial \boldsymbol{\zeta} }\left(\frac{\partial \ln(p_Z(\boldsymbol{\zeta} ))}{\partial \boldsymbol{\zeta} }\right)^H\right].
	\end{equation}
	
	In the following, we derive more tractable expressions of the FIMs in (\ref{Fo}) and (\ref{Fp}). First, for $\bm{F}_o$, the log-likelihood function for estimating $\bm{\zeta}$ from the observation $\bm{Y}$ is expressed as\cite{6}:
	\begin{align}
		\ln(f(\bm{Y}|\boldsymbol{\zeta}))=&\frac{2}{\sigma^2}\mathrm{Re}\{ \mathrm{tr}(\bm{X}^H\bm{G}^H(\theta)\bm{Y })\}\nonumber\\
		&-\frac{\|\bm{Y }\|_F^2 + \|\bm{G}(\theta)\bm{X}\|_F^2 }{\sigma^2}-N_rL\ln(\pi\sigma^2).
	\end{align}
	Since $\bm{G}(\theta)$ is a function of $\bm{a}(\theta)$ and $\bm{b}(\theta)$, $\bm{F}_o$ is a function of the derivatives of $\bm{a}(\theta)$ and $\bm{b}(\theta)$ denoted by $\dot{\bm{a}}(\theta)$ and $\dot{\bm{b}}(\theta)$, respectively, with $\dot{a}_i(\theta)=\frac{-j\pi d(N_t-2i+1)\cos\theta}{\lambda}a_i(\theta), i\!=\!1,...,N_t$ and $\dot{b}_m(\theta)=\frac{-j\pi d(N_r-2m+1)\cos\theta}{\lambda}b_m(\theta), m\!=\!1,...,N_r$. Note that $\boldsymbol{a}^H(\theta)\dot{\bm{a}}(\theta)=0$ and $\boldsymbol{b}^H(\theta)\dot{\bm{b}}(\theta)=0$. By leveraging this property, the FIM $\bm{F}_o$ in (\ref{Fo}) can be derived as \cite{estimationtheory}:
	\begin{align}
		\bm{F}_o
		= \left[
		\begin{array}{ll}
			J_{\theta\theta}            & \bm{J}_{\theta\alpha}     \\
			\bm{J}_{\theta\alpha}^H & \bm{J}_{\alpha\alpha}
		\end{array}
		\right].
	\end{align}
	Specifically, $J_{\theta\theta}$, $\bm{J}_{\theta\alpha}$, and $\bm{J}_{\alpha\alpha}$ are given by
	\begin{align}
		J_{\theta\theta}=&\frac{2|\alpha|^2L}{\sigma^2}\mathrm{tr}\left(\bm{A}_1\bm{R}_X\right)+\frac{2|\alpha|^2L N_r}{\sigma^2}\mathrm{tr}\left(\bm{A}_2\bm{R}_X\right),  \\
		\bm{J}_{\theta \alpha}=& \frac{2L N_r}{\sigma^2} \mathrm{tr}\left(\bm{A}_3\bm{R}_X\right)[\alpha_R,\alpha_I],\\
		\bm{J}_{\alpha\alpha}=& \frac{2L N_r}{\sigma^2}\mathrm{tr}\left(\bm{A}_4\bm{R}_X\right)  \bm{I}_{2},
	\end{align}
	where $\bm{A}_1=\int \|\dot{\bm{b}}(\theta)\|^2\bm{a}(\theta)\bm{a}^H(\theta) p_\Theta(\theta)d\theta $,  $\bm{A}_2=\int \dot{\bm{a}}(\theta)\dot{\bm{a}}^H(\theta) p_\Theta(\theta)d\theta $,  $\bm{A}_3=\int \dot{\bm{a}}(\theta)\bm{a}^H(\theta) p_\Theta(\theta)d\theta $, and $\bm{A}_4=\int \bm{a}(\theta)\bm{a}^H(\theta) p_\Theta(\theta)d\theta$.
	
	On the other hand, for $\bm{F}_p$, since $\alpha_R$ and $\alpha_I$ are deterministic parameters, we have $\frac{\partial
		\ln(p_Z(\boldsymbol{\zeta} ))}{\partial \boldsymbol{\zeta} }=\left[ \frac{\partial \ln(p_\Theta(\theta))}{\partial \theta},0,0\right]^T$, which yields $[\bm{F}_p]_{1,1}=\mathbb{E}_\theta\left[\left(\frac{\partial \ln(p_\Theta(\theta))}{\partial \theta}\right)^2\right]$ and $[\bm{F}_p]_{m,n}=0$ for any $(m,n)\neq (1,1)$. Let $f_{k}(\theta)=\frac{1}{\sqrt{2\pi}\sigma_k} e^{-\frac{(\theta-\theta_k)^2}{2\sigma_k^2}}$ denote each $k$-th Gaussian PDF in the Gaussian mixture model. Then, $[\bm{F}_p]_{1,1}$ can be expressed as
	\begin{align}\label{Fp11}
		&[\bm{F}_p]_{1,1}=\int \left(\frac{\partial \ln(p_\Theta(\theta))}{\partial \theta}\right)^2\!\!p_\Theta(\theta)d\theta  \!\! \\
		&=\!\sum\limits_{k=1}^{K} \frac{p_k}{\sigma_k^2} \!-\!\!\!\underbrace{  \mathlarger{\int} \frac{\sum\limits_{k_1=1}^{K}\!\sum\limits_{k_2=1 }^{K}\!p_{k_1}\!p_{k_2}\!f_{k_1}\!(\theta)\!f_{k_2}\!(\theta)\!\left(\!\!\frac{\theta-\theta_{k_1}}{\sigma_{k_1}^2}\!\!-\!\!\frac{\theta-\theta_{k_2}}{\sigma_{k_2}^2} \!\!\right)^2  }{ 2\sum\limits_{k=1}^{K}p_kf_k(\theta)}   d\theta}_\rho\!. \nonumber
	\end{align}
	Note that $[\bm{F}_p]_{1,1}=\sum_{k=1}^K \frac{p_k}{\sigma_k^2}-\rho\geq 0$ holds according to (\ref{Fp11}).
	
	Therefore, the overall FIM for $\boldsymbol{\zeta}$ is given by
	\begin{align}
		\boldsymbol{F} =\bm{F}_o +\bm{F}_p= \left[
		\begin{array}{cc}
			J_{\theta\theta}+ \sum\limits_{k=1}^{K}\frac{p_k}{\sigma_k^2}-\rho            & \bm{J}_{\theta\alpha}     \\
			\bm{J}_{\theta\alpha}^H & \bm{J}_{\alpha\alpha}
		\end{array}
		\right].
	\end{align}
	Note that the overall PCRB for estimating $\bm{\zeta}$ is determined by $\bm{F}^{-1}$ expressed as
	\begin{align}
		\bm{F}^{-1}
		= \left[
		\begin{array}{ll}
			S^{-1} & \bm{D}       \\
			\bm{D}^H           & \bm{E}
		\end{array}
		\right],
	\end{align}
	where $S \in \mathbb{C}$, $\bm{D} \in \mathbb{C}^{1\times2}$, and $\bm{E} \in \mathbb{C}^{2\times2}$. Particularly, $S$ is the Schur complement of block $ \bm{J}_{\alpha\alpha}$, which is given by
	\begin{align}\label{Fx}
		S \overset{\Delta}{=}
		J_{\theta\theta}+ \sum_{k=1}^{K}\frac{p_k}{\sigma_k^2}-\rho -  \bm{J}_{\theta\alpha} \bm{J}_{\alpha\alpha}^{-1} \bm{J}^H_{\theta\alpha} .
	\end{align}
	
	In this paper, we aim to derive the PCRB for estimating the target's angle $\theta$, which is only dependent on $S$, as given below:
	\begin{align}\label{PCRB}
		\mathrm{PCRB}_{\theta}&=[\bm{F}^{-1}]_{1,1}=S^{-1}\nonumber\\
		&=\frac{\sigma^2}{2|\alpha|^2L}\Bigg/\Bigg(\frac{\sigma^2}{2|\alpha|^2L}\left(\sum_{k=1}^{K}\frac{p_k}{\sigma_k^2}-\rho\right) + \mathrm{tr}\left( \bm{A}_1\bm{R}_X\right)\nonumber\\
		&+N_r  \mathrm{tr}\left(\bm{A}_2\bm{R}_X\right)-\frac{N_r\left| \mathrm{tr}\left(\bm{A}_3\bm{R}_X\right)\right|^2}{   \mathrm{tr}\left(\bm{A}_4\bm{R}_X\right)}\Bigg).
	\end{align}
	
	\subsection{Tractable Bound of PCRB}
	Note that the exact PCRB in (\ref{PCRB}) has a complicated expression, which is difficult to analyze and draw insights from; moreover, it can be shown to be a non-convex function over the sample covariance matrix $\bm{R}_X$ of the transmit signal, which makes it difficult to be used as an optimization objective function. To overcome these challenges, we propose an upper bound of the exact PCRB $\mathrm{PCRB}_{\theta}$, whose tightness will be verified numerically in Section \ref{sec_num}.
	\begin{proposition}\label{prop_bound}
		$\mathrm{PCRB}_{\theta}$ is  upper bounded as
		\begin{align}\label{bound}
			\mathrm{PCRB}_{\theta}\leq \mathrm{PCRB}_\theta^U\overset{\Delta}{=} \frac{\frac{\sigma^2}{2|\alpha|^2L}}{ \frac{\sigma^2}{2|\alpha|^2L}\left(\sum\limits_{k=1}^{K}\frac{p_k}{\sigma_k^2}-\rho\right)+\mathrm{tr}\left(\bm{A}_1\bm{R}_X\right)}.
		\end{align}
	\end{proposition}
	\begin{IEEEproof}
		Please refer to Appendix A.
	\end{IEEEproof}
	
	Notice that the PCRB upper bound in (\ref{bound}) is in a much simpler form compared to the exact PCRB in (\ref{PCRB}). In the following, we will leverage this upper bound for discussing the effect of exploiting prior information in the estimation of $\theta$, and for optimizing the sample covariance matrix $\bm{R}_X$.
	\subsection{Effect of Exploiting Prior Information}
	In this subsection, we aim to investigate the effect of exploiting prior distribution information on the estimation performance. Specifically, when the prior distribution information of $\theta$ is unknown, CRB can be adopted to characterize a lower bound of the estimation MSE corresponding to each realization of $\theta$, which is given as
	\begin{align}\label{CRB}
		\mathrm{CRB}_{\theta}(\theta)\!=\!\frac{1}{J_{\theta\theta}\!-\!\bm{J}_{\theta\alpha}\bm{J}_{\alpha\alpha}^{-1}\bm{J}_{\theta\alpha}^H}
		\!=\!\frac{\frac{\sigma^2}{2|\alpha|^2L }}{ \|\dot{\bm{b}}(\theta)\|^2 \mathrm{tr}\left(\bm{a}(\theta)\bm{a}^H(\theta)\bm{R}_X\right)}.
	\end{align}
	Moreover, by taking the expectation of $\mathrm{CRB}_{\theta}(\theta)$ over the random angle realizations, the average (expected) CRB is given by
	\begin{align}
		\mathrm{CRB}_{\theta} =\mathbb{E}_{\theta}[\mathrm{CRB}_{\theta}(\theta)]=\int \mathrm{CRB}_{\theta}(\theta)   p_\Theta(\theta)d\theta,
	\end{align}
	which can be viewed as a lower bound of the long-term MSE performance without exploiting prior information. Note that based on Jensen's inequality and $\sum_{k=1}^K \frac{p_k}{\sigma_k^2}-\rho\geq 0$, we have
	\begin{align}
		\mathrm{CRB}_{\theta}=&\mathbb{E}_{\theta}\left[\frac{\frac{\sigma^2}{2|\alpha|^2L}}{\|\dot{\bm{b}}(\theta)\|^2\mathrm{tr}\left(\bm{a}(\theta)\bm{a}^H(\theta)\bm{R}_X\right)}\right]\nonumber\\
		\geq&\frac{1}{\mathbb{E}_{\theta}\left[\frac{\|\dot{\bm{b}}(\theta)\|^2\mathrm{tr}\left(\bm{a}(\theta)\bm{a}^H(\theta)\bm{R}_X\right)}{\frac{\sigma^2}{2|\alpha|^2L}}\right]}=\frac{\frac{\sigma^2}{2|\alpha|^2L}}{ \mathrm{tr}\left(\bm{A}_1\bm{R}_X\right)}\nonumber\\
		\geq&\mathrm{PCRB}_\theta^U\geq \mathrm{PCRB}_{\theta}.
	\end{align}
	
	The above result indicates that exploiting the prior distribution information can achieve a decreased lower bound on the estimation MSE. Since the CRB/PCRB is generally tight in the moderate-to-high signal-to-noise ratio (SNR) regime, this further implies that the estimation performance can be improved via the exploitation of prior information. Moreover, according to the properties of Jensen's inequality \cite{gap}, the gap between $\mathrm{CRB}_{\theta}$ and $\mathrm{PCRB}_{\theta}$ generally increases as the variance of $\theta$ increases. Hence, for scenarios where the possible target locations are more dispersed, the performance gain via exploiting prior distribution information will be more significant.

	\section{Problem Formulation}\label{sec_pro}
	In this section, we formulate the problem of optimizing the sample covariance matrix of the MIMO radar transmit signals for estimation performance optimization exploiting prior distribution information. Specifically, since the exact PCRB is in a complicated form, we aim to minimize the upper bound of PCRB derived in (\ref{bound}), subject to a total power constraint at the MIMO radar transmit antennas. The optimization problem is formulated as
	\begin{align} \label{P1}
		(\mathrm{P1})\quad  \mathop{\mathtt{minimize}}_{\bm{R}_X} \quad & \mathrm{PCRB}^U_{\theta}
		\\
		\mathtt{subject}\;\mathtt{to} \quad & \mathrm{tr}\left(\bm{R}_X\right)\leq P, \\
		& \bm{R}_X\succeq {\bm{0}}.
	\end{align}
	
	Note that $\bm{R}_X$ only affects the denominator in $\mathrm{PCRB}_{\theta}^U$ shown in (\ref{bound}), the maximization of which is equivalent to the maximization of $\mathrm{tr}(\bm{A}_1\bm{R}_X)$. Hence, Problem (P1) is equivalent to Problem (P2), which is given by:
	\begin{align} \label{P2}
		(\mathrm{P2})\quad  \mathop{\mathtt{maximize}}_{\bm{R}_X} \quad & \mathrm{tr}(\bm{A}_1\bm{R}_X)
		\\
		\mathtt{subject}\;\mathtt{to} \quad & \mathrm{tr}\left(\bm{R}_X\right)\leq P, \\
		& \bm{R}_X\succeq {\bm{0}}.
	\end{align}
	
	In the following, we will obtain the optimal solution to (P1) via solving (P2).
	\section{Optimal Solution}\label{sec_solution}
	Problem (P2) is a semi-definite program (SDP), for which the optimal solution can be obtained via the interior-point method or existing software, e.g., CVX. To draw more useful insights, we present the optimal solution in closed form.
	
	First, since there is only one linear constraint in (P2), there exists a rank-one optimal solution to (P2) denoted by $\bm{R}^\star_X=\bm{w}^\star\bm{w}^{\star H}$ \cite{Palomar}. This implies that using a constant probing signal $\bm{w}^\star$ for all the samples is already optimal. Consequently, (P2) can be equivalently transformed into the following problem for optimizing the constant probing signal denoted by $\bm{w}\in \mathbb{C}^{N_t\times 1}$:
	\begin{align} \label{P3}
		(\mathrm{P3})\quad  \mathop{\mathtt{maximize}}_{\bm{w}} \quad & \bm{w}^H\bm{A}_1\bm{w}
		\\
		\mathtt{subject}\;\mathtt{to} \quad & \|\bm{w}\|^2\leq P.
	\end{align}
	
	To solve (P3), we express the eigenvalue decomposition (EVD) of $\bm{A}_1$ as $\bm{A}_1=\bm{Q}\bm{\Lambda}\bm{Q}^H$, where $ \bm{\Lambda}=\mathrm{diag}\{\lambda_1,...,\lambda_{N_t}\}$, with $\lambda_1\geq \lambda_2\geq ...\geq \lambda_{N_t}\geq 0$; $\bm{Q}=[\bm{q}_1,...,\bm{q}_{N_t}]$ is a unitary matrix with $\bm{Q}\bm{Q}^H=\bm{Q}^H\bm{Q}=\bm{I}_{N_t}$. Define $\boldsymbol{g}\triangleq \bm{Q}^H\boldsymbol{w}$. The objective function of (P3) can be further rewritten as $
	\bm{w}^H\bm{A}_1\bm{w}= \boldsymbol{w}^H\bm{Q}\bm{\Lambda}\bm{Q}^H\boldsymbol{w}
	=\boldsymbol{g}^H \bm{\Lambda}\boldsymbol{g}$. The constraint can be represented as $\|\bm{w}\|^2=\|\bm{g}\|^2\leq P$. Hence, (P3) can be equivalently transformed into the following problem:
	\begin{align} \label{P4}
		(\mathrm{P4})\quad  \mathop{\mathtt{maximize}}_{\bm{g}} \quad & \boldsymbol{g}^H \bm{\Lambda}\boldsymbol{g}
		\\
		\mathtt{subject}\;\mathtt{to} \quad & \|\bm{g}\|^2 \leq P.
	\end{align}
	Note that for any feasible solution $\bm{w}$ to (P3), $\bm{g}=\bm{Q}^H\bm{w}$ is a feasible solution to (P4) with the same objective value; while for any feasible solution $\bm{g}$ to (P4), $\bm{w}=\bm{Q}\bm{g}$ is a feasible solution to (P3) with the same objective value. Thus, (P4) is equivalent to (P3) and consequently (P2) and (P1).
	
	Based on (P4), we have the following proposition.
	\begin{proposition}\label{prop_2}
		An optimal solution to (P1) is $\boldsymbol{R}^\star_X= P\bm{q}_1\bm{q}_1^H$.
	\end{proposition}
	\begin{IEEEproof}
		The objective value of (P4) is upper bounded as
		\begin{align}
			\boldsymbol{g}^H \bm{\Lambda}\boldsymbol{g}=\sum_{i=1}^{N_t}|g_i|^2\lambda_i\leq \sum_{i=1}^{N_t}|g_i|^2 \lambda_1\leq P \lambda_1.
		\end{align}
		The above equalities hold when $\bm{g}=[\sqrt{P},0,...,0]^T$, which is thus the optimal solution to (P4). Hence, the optimal constant probing signal in (P3) is given by $\bm{w}^\star=\bm{Q}\bm{g}=\sqrt{P}\bm{q}_1$, where $\bm{q}_1$ is the eigenvector corresponding to the largest eigenvalue (i.e., $\lambda_1$) of $\bm{A}_1$. Thus, $\boldsymbol{R}^\star_X= P\bm{q}_1\bm{q}_1^H$ is optimal to (P1).
	\end{IEEEproof}
	
	Note that the complexity for obtaining the optimal solution to (P1) via Proposition \ref{prop_2} can be shown to be $O(N_t^3)$ \cite{SVD}, which is lower than that via the interior-point method for SDP, i.e., $\mathcal{O}(N_t^{7})$ \cite{SDR4}.
	
	Based on Proposition \ref{prop_2}, the minimum value of the PCRB upper bound with optimized transmit signals is obtained as
	\begin{align}
		\mathrm{PCRB}_\theta^{U^\star}
		=\frac{1}{\sum\limits_{k=1}^{K}\frac{p_k}{\sigma_k^2}-\rho+\frac{2P|\alpha|^2L}{\sigma^2}\bm{q}_1^H\bm{A}_1\bm{q}_1}.
	\end{align}
	It is worth noting that the PCRB upper bound decreases as the term $\frac{P|\alpha|^2L}{\sigma^2}$ increases, which denotes the overall SNR at the MIMO radar receiver.
	
	\section{Numerical Results}\label{sec_num}
	In this section, we provide numerical results to evaluate the performance of the proposed MIMO radar transmit signal design. We consider a MIMO radar system with $N_t=10$ transmit antennas and $N_r=12$ receive antennas, where the antenna spacing is set as $d=\frac{\lambda}{2}$. The number of signal samples is set as $L=25$. The transmit power is set as $P=30$ dBm. The average noise power is set as $\sigma^2=-120$ dBm. For the PDF of angle $\theta$, we set $K=5$; $\theta_1=0.52 $, $\theta_2=0.82 $, $\theta_3=0.87 $, $\theta_4=2.6 $, $\theta_5=2.7 $; $\sigma_1^2=10^{-4}$, $\sigma_2^2=10^{-4}$, $\sigma_3^2=10^{-3}$, $\sigma_4^2=10^{-3}$, $\sigma_5^2=10^{-4}$; $p_1=0.15$, $p_2=0.32$, $p_3=0.17$, $p_4=0.2$, $p_5=0.16$. For comparison, we consider the following benchmark schemes for the transmit signal design:
	\begin{itemize}
		\item {\bf{Benchmark Scheme 1: Heuristic signal design}}. In this scheme, we design $\bm{R}_X$ as a diagonal matrix with signal power allocated to the first transmit antenna, i.e., $\bm{R}_X=\mathrm{diag}\{P,0,...,0\}$ and $\bm{x}_l=[\sqrt{P},0,...,0]^T,\forall l$.
		\item {\bf{Benchmark Scheme 2: Highest-probability angle based signal design}}. In this scheme, we design $\bm{R}_X$ to minimize the CRB in (\ref{CRB}) corresponding to the angle with highest probability, i.e., $\mathrm{CRB}_{\theta}({\theta}_{\max})$ where ${\theta}_{\max}=\arg\max\ p_\Theta(\theta)$. It can be shown in a similar manner as that in Section \ref{sec_solution} that the optimal solution is $\boldsymbol{R}_X=\frac{P}{N_t}\boldsymbol{a}({\theta}_{\max})\boldsymbol{a}^H({\theta}_{\max})$ and   $\bm{x}_l=\sqrt{\frac{P}{N_t}}\boldsymbol{a}({\theta}_{\max}),\forall l$.
	\end{itemize}

\begin{figure}[t]
		\centering
		\includegraphics[width=3.5in]{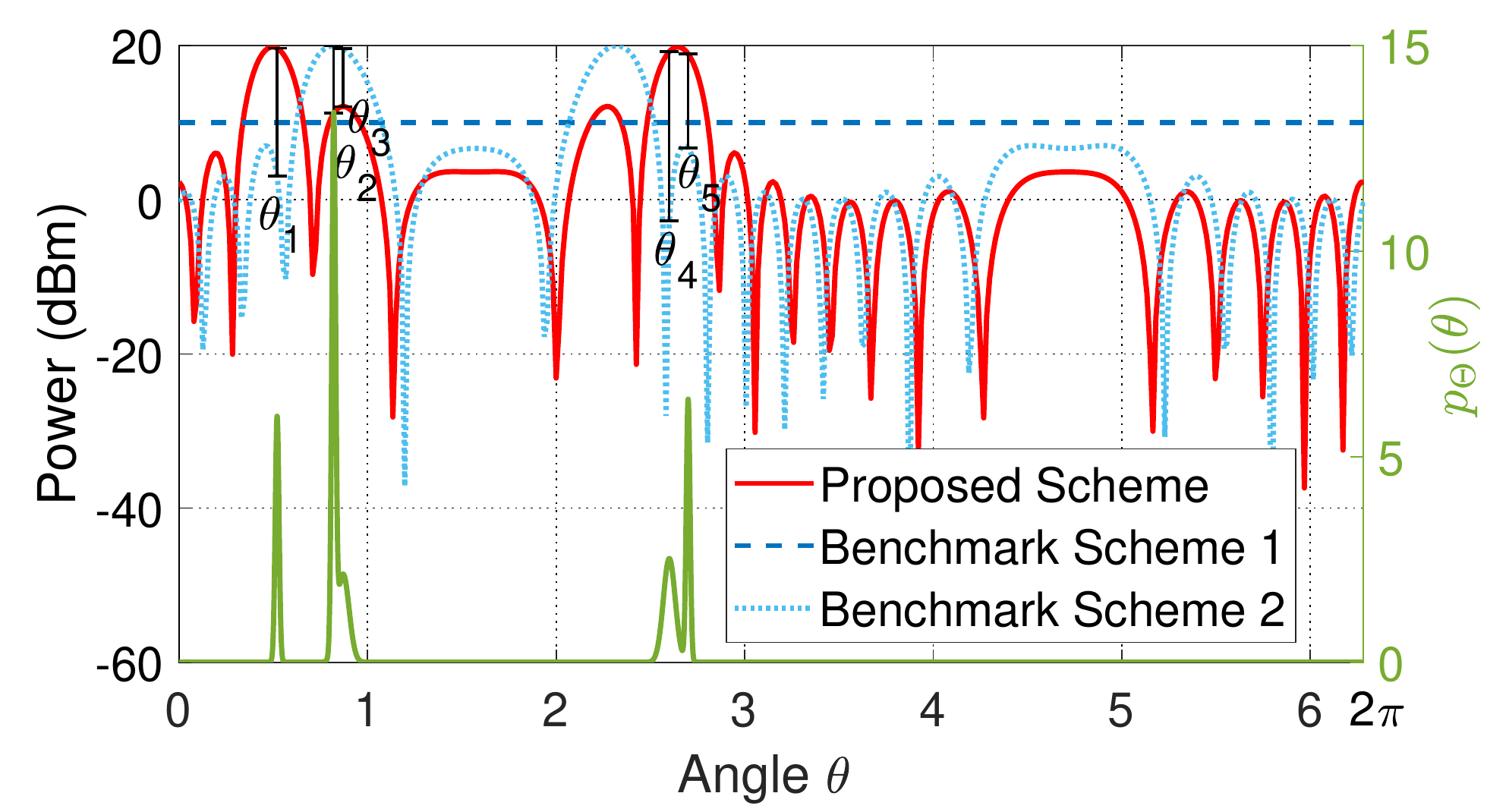}
		\vspace{-8mm}
		\caption{Radiated power pattern and $p_\Theta(\theta)$ over different angles.}\label{Fig_power}
		\vspace{-3mm}	\end{figure}

	First, we show in Fig. \ref{Fig_power} the radiated power pattern at distance $r$ with different transmit signal designs and the prior PDF of $\theta$ over different angles. We set $\frac{\beta_0}{r^2}=-20$ dB. It is observed that the proposed scheme achieves higher radiation power at all the angles with non-zero probability densities compared to Benchmark Scheme 1, since the latter yields an omni-directional radiation pattern with the power wasted on the angles with zero probability density. Moreover, the proposed scheme outperforms Benchmark Scheme 2 for most angles with non-zero probability densities, at the cost of only a small power loss around the highest-probability angle. This shows that the proposed scheme is able to achieve a more balanced power distribution over the possible angles, by judiciously designing the transmit signals based on the prior distribution information.
	
	\begin{figure}[t]
		\centering
		\includegraphics[width=3.4in]{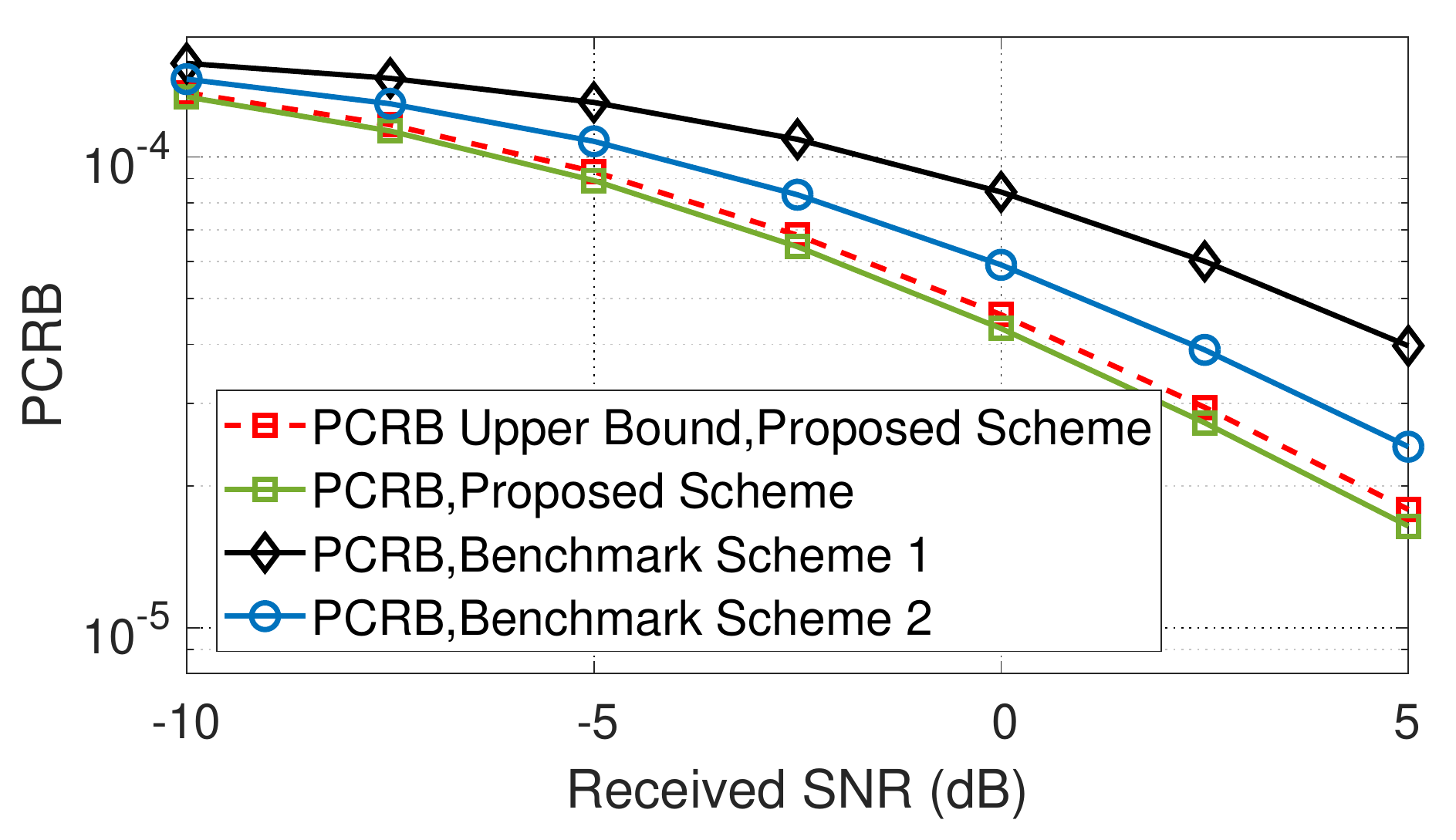}
		\vspace{-4mm}
		\caption{PCRB versus received SNR.}\label{Fig_PCRB}
		\vspace{-7mm}	\end{figure}
	
	Next, in Fig. \ref{Fig_PCRB}, we evaluate the PCRB achieved by different transmit signal design schemes. We also show the PCRB upper bound achieved via our proposed scheme. It is observed that our proposed PCRB upper bound is tight for all SNR regimes, which verifies the effectiveness of using this upper bound as the estimation performance metric. Moreover, it is observed that the proposed scheme outperforms both Benchmark Scheme 1 and Benchmark Scheme 2, due to the smart exploitation of the prior distribution information.

	\section{Conclusions}
	Considering a MIMO radar system, this paper studied the angle estimation of a target exploiting its prior distribution information. The PCRB of the angle estimation MSE was first derived, for which a more tractable and tight upper bound was proposed. It was analytically shown that by exploiting the prior information, the PCRB is guaranteed to be no larger than the average CRB without exploiting prior information. Next, the problem of transmit signal optimization was formulated, with the objective of minimizing the PCRB upper bound. The optimal sample covariance matrix was revealed to have a rank-one structure, based on which the optimal solution was derived in closed form. It was shown via numerical results that the proposed transmit signal design significantly outperforms various benchmark schemes.

		\appendices
		\section{Proof of Proposition \ref{prop_bound}}\label{proof_bound}
		By noting that the sample covariance matrix is given by $\bm{R}_X=\frac{1}{L}\sum_{l=1}^{L}\bm{x}_l\bm{x}_l^H$, we have the following inequality:
		\begin{align}\label{inequality}
			&\left(\mathrm{tr}\left(\bm{A}_2\bm{R}_X\right) \mathrm{tr}\left(\bm{A}_4\bm{R}_X\right)- \left| \mathrm{tr}\left(\bm{A}_3\bm{R}_X\right)\right|^2\right)L^2 \nonumber\\
			=&\int \sum\limits_{l=1}^{L}|\boldsymbol{\dot{a}}^H(\theta)\boldsymbol{x}_l|^2 p_\Theta(\theta)d\theta\int \sum\limits_{l=1}^{L}|\boldsymbol{a}^H(\theta)\boldsymbol{x}_l|^2 p_\Theta(\theta)d\theta \nonumber\\
			&-\left|\int \sum\limits_{l=1}^{L}\boldsymbol{a}^H(\theta)\boldsymbol{x}_l\boldsymbol{x}_l^H\boldsymbol{\dot{a}}(\theta) p_\Theta(\theta)d\theta\right|^2 \nonumber\\
			=&\int\!\!\int \Bigg( \!\!-\!\left( \sum\limits_{l=1}^{L}\boldsymbol{a}^H(\theta_a)\boldsymbol{x}_l\boldsymbol{x}_l^H\boldsymbol{\dot{a}}(\theta_a) \right)\!\!\left( \sum\limits_{l=1}^{L}\boldsymbol{a}^H(\theta_b)\boldsymbol{x}_l\boldsymbol{x}_l^H\boldsymbol{\dot{a}}(\theta_b)\!\! \right)\nonumber\\
			&+\frac{1}{2}\left(\sum\limits_{l=1}^{L}|\boldsymbol{\dot{a}}^H(\theta_a)\boldsymbol{x}_l|^2 \right)\!\!\left(\sum\limits_{l=1}^{L}|\boldsymbol{ a}^H(\theta_b)\boldsymbol{x}_l|^2\right) \nonumber\\
			&+\frac{1}{2} \left(\!\sum\limits_{l=1}^{L}|\boldsymbol{ a}^H\!(\theta_a)\boldsymbol{x}_l|^2 \!\!\right)\!\!\left(\!\sum\limits_{l=1}^{L}|\boldsymbol{\dot{a}}^H\!(\theta_b)\boldsymbol{x}_l|^2 \!\!\right)\!\! \Bigg)\!p_\Theta\!(\theta_a)p_\Theta\!(\theta_b)d\theta_a\! d\theta_b \nonumber\\
			=&\int\int \sum\limits_{n=1}^{L}\sum\limits_{l=1}^{L}\Bigg(\frac{1}{2} \left|\boldsymbol{\dot{a}}^H(\theta_a)\bm{x}_n\bm{x}_l\bm{a}^H(\theta_b)\right|^2 \nonumber\\
			&- \left(\boldsymbol{\dot{a}}^H(\theta_a)\bm{x}_n\bm{x}_l\bm{a}^H(\theta_b)\right)\left(\boldsymbol{{a}}^H(\theta_a)\bm{x}_n\bm{x}_l\bm{\dot{a}}^H(\theta_b) \right)\nonumber\\
			&+\frac{1}{2} \left|\boldsymbol{{a}}^H(\theta_a)\bm{x}_n\bm{x}_l\bm{\dot{a}}^H(\theta_b)\right|^2 \Bigg)p_\Theta(\theta_a)p_\Theta(\theta_b)d\theta_a d\theta_b \nonumber\\
			=&\int\int\sum\limits_{n=1}^{L}\sum\limits_{l=1}^{L}\frac{1}{2}\left|\boldsymbol{\dot{a}}^H(\theta_a)\bm{x}_n\bm{x}_l\bm{a}^H(\theta_b)\right.\nonumber\\
			&-\left.\boldsymbol{{a}}^H(\theta_a)\bm{x}_n\bm{x}_l\bm{\dot{a}}^H(\theta_b)\right|^2p_\Theta(\theta_a)p_\Theta(\theta_b)d\theta_a d\theta_b \nonumber\\
			\geq&0.
		\end{align}
		By applying (\ref{inequality}) on the denominator of (\ref{PCRB}), the proof of Proposition \ref{prop_bound} is completed.
		
	\bibliographystyle{IEEEtran}
	\bibliography{reference}
	
\end{document}